# Excess Noise in High-Current Diamond Diodes – Physical Mechanisms and Implications for Reliability Assessment

Subhajit Ghosh[1], Harshad Surdi[2], Fariborz Kargar[1], Franz A. Koeck[3], Sergey Rumyantsev[4], Stephen Goodnick[2,4], Robert Nemanich[3] and Alexander A. Balandin[1,×]

[1]Department of Electrical and Computer Engineering, Bourns College of Engineering, University of California, Riverside, California 92521 USA

[2]School of Electrical, Computer and Energy Engineering, Arizona State University, Tempe, Arizona 85281 USA

[3]Department of Physics, Arizona State University, Tempe, Arizona 85281 USA

[4]CENTERA Laboratories, Institute of High-Pressure Physics, Polish Academy of Sciences, Warsaw 01-142 Poland

**Abstract**

We report results of an investigation of low-frequency excess noise in high-current diamond diodes. It was found that the electronic excess noise of the diamond diodes is dominated by generation – recombination noise, which reveals itself either as Lorentzian spectral features or as a $1/f$ noise spectrum ($f$ is the frequency). The generation – recombination bulges are characteristic for diamond diodes with lower turn-on voltages. The noise spectral density dependence on forward current, $I$, reveals three distinctive regions in all examined devices – it scales as $I^2$ at the low ($I<10$ μA) and high ($I>10$ mA) currents, and, rather unusually, remain nearly constant at the intermediate current range. The characteristic trap time constants, extracted from the noise data, reveal a uniquely strong dependence on current. Interestingly, the performance of the diamond diodes improves with increasing temperature. The obtained results are important for development of noise spectroscopy-based approaches for device reliability assessment for the high-power diamond electronics.

×Corresponding author (AAB); E-mail: balandin@ece.ucr.edu ; https://balandingroup.ucr.edu/

Ultra-wide bandgap (UWBG) semiconductors have emerged as important materials for power converters to meet the increasing efficiency needs [1-6]. Diamond is promising UWBG material in terms of its critical electric field, drift velocity, carrier mobility, and thermal conductivity [7-13]. However, diamond chemical vapor deposition growth, processing, and doping are still in the early stages of development. Diamond technology is not as mature as that of GaN or SiC. Rather large concentrations of defects, including trap levels within the bandgap, can detrimentally affect diamond diode operation, *e.g.* by an increase in the diode turn-on voltage [8, 14-18]. Typical traps in diamonds have energy levels ranging from 0.2 $eV$ to 1.7 $eV$. In the $n$-layer of the diode, the P dopant activation energy level is 0.43 $eV$ to 0.63 $eV$ [19,20]. Due to the high activation energies, only a small fraction of the dopant atoms is ionized. The defects and impurities, acting as charge carrier traps, negatively affect the reliability of the diodes, which is one of the most important metrics for applications in power converters for electricity grids. These considerations explain the need for developing innovative techniques for assessing the material quality and reliability of diamond diodes.

Low-frequency electronic noise, also referred to as excess noise, has been used in characterizing reliability-limiting defects and impurities in electronic materials and devices [21-25]. The excess noise includes $1/f$ and generation – recombination (G-R) noise with a Lorentzian type spectrum, which add up to the thermal and shot noise background ($f$ is the frequency). It is known that $1/f$ noise can be an early indicator of electromigration damage, and provide insight into the nature and energy distributions of reliability-limiting defects in the as-processed and aged materials and devices [21,26-29]. The excess noise often originates in the non-ideal components or the non-ideal currents of a device. These include leakage current, defects in the material, or parasitic resistances. The noise level increases at a much faster rate than the DC parameters as a device degrades under stress or as a result of its aging [30]. For this reason, noise can be used as a sensitive predictor of lifetime. The rate of increase of the noise level can be related to the device mean-time-to-failure (MTTF). Since only short-time noise measurements are needed, the procedure for determining MTTF is non-destructive. In order to develop noise-based reliability assessment techniques for diamond diodes, one needs to conduct thorough studies of excess noise for this specific device type. The currently available data on noise in diamond devices are limited, with only a few reports published [31,32].

In this Letter, we report the results of the investigation of excess noise in diamond diodes designed specifically for applications as high-current switches. The main objective of the study is to provide the noise-level baseline required for the development of reliability assessment. The diamond diodes were grown on a <111> highly B doped ($\sim 2 \times 10^{20}$/cm$^3$) single crystal diamond plate (3×3×0.3 mm; TISNCM). A ~0.2 μm i-layer was grown on top of the B doped p$^{++}$ substrate using a plasma enhanced chemical vapor deposition (PECVD) with a mixture of H$_2$:CH$_4$:O$_2$ under a chamber pressure of 63 Torr and at 1000 W of microwave power. A ~0.15 μm moderately P doped (~10$^{18}$/cm$^3$) n-layer was grown on top of the i-layer using a H$_2$:CH$_4$:TMP mixture under 60 Torr pressure and 2000 W microwave power in the PECVD chamber. A ~0.1 μm near-metallic highly conductive N doped nano-carbon (nanoC) layer was grown on top of the n-layer to lower the contact resistance of the cathode contact. Additional growth details can be found in Ref. [33]. The active area of the diodes was defined by partially mesa etching the diamond into the i-layer using a SiO$_2$ hard mask and O$_2$/SF$_6$ chemistry in a reactive ion etcher. The top cathode and bottom anode contacts were defined by UV photolithography and e-beam deposition of a Ti-Ni-Au metal stack with 50 nm - 50 nm – 300 nm thicknesses. The layered structure of the devices and band diagram are shown in Figure 1 (a-b).

[Figure 1: (a) Schematic of the layered structure of diamond diodes. (b) The band diagram of a diamond diode including the trap levels as simulated by Silvaco ATLAS.]

Although the nanoC layer provides a reduced contact resistance at the circular cathode contact, a Schottky barrier height (SBH) to electrons exists due to the Fermi-level pinning at the nanoC-n-layer interface. The SBH at the cathode contact can be seen from the band diagram generated from Silvaco ATLAS simulations using the diode structure (see Figure 1 (a-b)). The diode current is therefore largely dominated by the holes injected over the p$^{++}$-i-n junction barrier. The hole injection initially follows thermionic emission (TE) regime where the current is exponentially dependent on the small forward bias voltage, and then transitions into a space charge limited conduction (SCLC) where the current is proportional to $V^m$ where m is ≥ 2 depending on the trap energy levels and trap density in the diode [8,34].

The current-voltage (*I-V*) and low-frequency noise characteristics for all diodes were measured in vacuum (Agilent; Lake Shore). The noise spectra were acquired with a dynamic signal analyzer (Stanford Research). Details of our noise measurements procedures, in the context of other material systems, have been reported by some of us elsewhere [35-37]. Figures 2 (a) and (b) show the *I-V* characteristics in forward bias for six different diamond diodes in linear and logarithmic scales, respectively. From the *I-V* characteristics, the tested diodes can be separated into two groups based on the respective turn-on voltage, $V_T$, of each diode. Devices 1, 2, and 3 have lower turn-on voltages, close to 5 V as compared to devices 4, 5, and 6 that have turn-on voltages of ~10 V or higher. The deviations from the ideal characteristics of these diodes and current jumps indicate the material imperfections owing to the infancy of the diamond diode growth and processing technology.

The noise measurements were conducted for all devices to examine variations in the noise characteristics for diodes with different turn-on voltages. In Figure 2 (c) we present the noise current spectral density, $S_I$, as a function of frequency for a device 2 ($V_T$~5 V) for different currents through the diode. The noise spectra at all currents are the superposition of the 1/*f* and G-R noise with the pronounced Lorentzian spectral features. Similar noise spectra were measured for devices 1 and 3 with the low turn-on voltage. The noise spectral density, $S_I$, as a function of frequency for a device 6 with the highest turn-on voltage is presented in Figure 2 (d). Interestingly, in this and other high turn-on voltage diodes, the noise is predominantly of the *1/f*-type. Some signatures of the deviation from 1/*f* spectrum and contribution of the G-R noise appear only at higher frequency (*f*~10 kHz). These noise characteristics were similar for all diodes with the high turn-on voltages. The diamond diodes with the larger tun-on voltages are typically those that have a higher concentration of defects [8]. One may consider it unusual that the devices with more defects have 1/*f* spectrum while those with fewer defects show G-R spectral features. Below we explain it by the specifics of the noise mechanisms in diodes as compared to linear resistors or field-effect transistors (FETs).

[**Figure 2:** Current-voltage characteristics of six different diamond diodes plotted in linear (a) and logarithmic (b) scales. The diodes can be grouped as those with lower ($V_T$~5 V) and higher ($V_T$~10 V) turn-on voltages. (c) Noise current spectral density, $S_I$, as a function of frequency for different current densities, *J*, for a device 2, with the low turn-on voltage. (d) Noise current spectral density, $S_I$, as a function of frequency for different current densities, $J$, for a device 6, with the high turn-on voltage.]

Figure 3 (a) shows the noise current spectral density, $S_I$, as a function of the current density, $J$, at a fixed frequency $f$=10 Hz for all six tested devices. The $S_I$ vs. $J$ ($S_I$ vs. $I$) relation for the measured diamond diodes follows the $S_I \sim J^2$ ($S_I \sim I^2$) trend at lower currents. Although the dependence $S_I \sim I$ can be considered as a typical one for forward biased diodes, $S_I \sim I^2$ dependences were also reported in *p-n* junction and Schottky diodes, including those based on wide-band-gap semiconductors [38]. The $S_I(J)$ dependence becomes almost flat in the intermediate current density range, *i.e.* from about $J\sim$0.1 A/cm$^2$ to 100 A/cm$^2$. The $S_I \sim J^2$ noise behavior is restored at higher currents. In is interesting to note that the transition to the flat $S_I(J)$ dependence at the intermediate current levels roughly corresponds to the transition from the TE to SCLC hole transport regimes. In previous studies, the non-monotonic trends in the noise spectral density of GaN diodes were interpreted as the interplay of contributions to noise from the diode base, *p-n* junction region, and the series resistances associated with the contacts [39]. A quantitative description for a non-monotonic trend was developed for SiC diodes [40].

To compare the noise characteristics of diamond diodes with those made of other material systems, we present the noise spectral density normalized by the current and device area in Figure 3 (b). For all measured diamond diodes, the normalized noise spectral density, $S_I/I^2 \times \Omega$, measured at fixed $f$=10 Hz, decreases with the increasing current density, $J$ ($\Omega$ is the area of the top cathode contact). The decrease becomes slow or saturates completely at high current densities when the contributions from series resistances start to dominate the noise response. A comparison of the noise spectral density normalized to the area indicates a substantially higher noise level in diamond than that in GaN diodes [25]. The *I-V* and noise characteristics of the diamond diodes at elevated temperatures are shown in Figures 3 (c) and (d). One can see that both current and noise are weak functions of temperature, which are beneficial for the high-power switching applications. Despite the high thermal conductivity of diamond, the high-current diodes can still experience substantial Joule heating at the considered power levels [8,41,42]. The thermal interface resistances between the layers increase the overall thermal resistance of the device structure [43]. The dashed lines in Figure 3 (c) show the slope of the exponential part of *I-V* characteristics, *e.g.*, $I\sim exp(V\eta KT)$, at low currents. The ideality factor, $\eta$, decreases with the temperature increase. The decrease in the ideality factor together with almost constant noise level with temperature indicates that the performance of the diamond diodes can be improved with increasing temperature. The latter is an extra benefit for applications of diamond diodes as high-power switches.

[**Figure 3:** (a) The noise current spectral density, $S_I$, as a function of the current density, $J$, at $f$ = 10 Hz for all devices, measured at room temperature. (b) The normalized noise current spectral density, $S_I/I^2 \times \Omega$, as a function of $J$ at $f$ = 10 Hz. (c) Current-voltage characteristics of a diamond diode (device 2) at elevated temperatures. (d) The noise spectral density, $S_I$, at $f$ = 10 Hz, as a function of temperature, measured for different current densities.]

We now turn to a more detailed analysis of the G-R bulges observed in the spectra of the low turn-on voltage devices. Figures 4 (a) and (b) show the noise spectra of device 1 for $J$ = 1.3 A/cm$^2$ and $J$ =1.3×10$^{-3}$ A/cm$^2$, respectively. One can see an overlap of two Lorentzians at the intermediate current levels (panel (a)), and just one pronounced Lorentzian superimposed over 1/$f$ shape at the small current level (panel (b)). We used fitting with Lorentzian to determine the corner frequency of the G-R noise. The G–R noise spectral density is described by the Lorentzian using an expression $S_I(f) = S_0/[1 + (2\pi f\tau)^2]$, where $S_0$ is the frequency independent portion of $S_I(f)$ observed at $f \ll (2\pi\tau)^{-1}$ and $\tau$ is the time constant associated with a particular fluctuation process. Figures 4 (c) and (d) show the characteristic frequency $f_c=(1/2\pi\tau)$ as a function of $J$ for diamond diodes 1 and 2, respectively. The general trend for $f_c$ is to increase with $J$. The functional dependence can be approximated as $f_c \sim J^\beta$, where $\beta$ takes the values 0.31 and 1.15 for device 1, 1.39 for the device 2 and 1.35 for device 3 [Refer to the supplemental figure S6]. The dependence of $f_c$ on current in the diodes can be explained by the dependence of the trap capture time $\tau_{cap}=(nv\sigma)^{-1}$ on the carriers concentration, $n$ (here $v$ is the charge carrier thermal velocity and $\sigma$ is the capture cross section) [44]. The current increase leads to the increase of the concentration, $n$, and corresponding increase of the corner frequency, $f_c,=(1/2\pi\tau)$. This is true if the time constant $\tau$ is dominated by capture rather than emission time. Since our diamond diodes are characterized by deeper, not fully ionized donor and acceptor states, high trap concentration, and the hole-dominated charge transport, the obtained $f_c(J)$ relations for the diamond diodes are substantially different from those reported for GaN diodes [44]. It is interesting to note that characteristic frequencies for some Lorentzian, *e.g.* red line in Figure 4 (c), coincide with the current jumps in *I-V* characteristics – see the current jump at 3.3 V in Figure 2 (a). This adds support to the model, which correlates current jumps in the I-Vs with the traps in the energy band gap [8].

[**Figure 4:** (a) The current noise spectral density, $S_I$, as a function of frequency at an intermediate current density, $J = 1.3$ A/cm$^2$, for a device 1 with the low turn-on voltage. (b) The current noise spectral density, $S_I$, as a function of frequency at a low current density, $J = 1.3 \times 10^{-2}$ A/cm$^2$, for the same device. Dependence of the corner frequency of the G-R bulges on the current density shown for (c) device 1 and (d) device 2.]

The G-R noise mechanisms in diodes are different from the G-R and 1/*f* noise mechanisms in bulk semiconductors and FETs [26,45]. In the McWhorter model for FETs, the 1/*f* noise emerges as an overlap of Lorentzian bulges due to traps with different time constants. The time constant, $\tau$, of the trap is determined by its distance from the conduction channel, *e.g.*, $\tau = \tau_0 exp(\lambda z)$, where $z$ is the distance of the trap from the channel, $\tau_0 \sim 10^{-10}$ s and $\lambda$ is the tunneling parameter. In bulk semiconductors, one needs several levels or a continuous band of trap levels with different capture cross sections, like density of state tails near conduction and valence band edges, in order to construct the 1/*f* noise spectrum [46]. The situation in diodes is different. One of the conventional models for the low-frequency noise in a diode assumes that the emission and capture of carriers by the recombination level leads to the fluctuations of the charge state of this trap, and, as a result, to the fluctuations of the electric field distribution in the space charge region with the corresponding current fluctuations [47]. This model has been refined to correlate the current fluctuations with the fluctuations of the electric field not in the entire space-charge region but rather in a small vicinity of a trap in a specific location in the p-n junction region [40]. The local fluctuations of the electric field are caused by the fluctuations of the charge state of the trap due to the exchange of electrons between this trap and the conduction band. Within this model, the same type of trap can give different time constants depending on its position in the diode structure since its energy level with respect to the Fermi energy is a strong function of the coordinate along the diode structure (see Figure 1 (b)). This can explain the evolution of G-R noise spectra to 1/*f* noise, and observed differences in the noise for the same type of devices that have low and high turn-on voltages (see Figure 2 (c-d)).

The diodes with the high turn-on voltage, which typically have more traps, have sufficient variation in the time constant, integrated over the junction length, to smooth out the G-R bulges to 1/$f$ envelope.

The noise in diamond diodes shows variations not only for the devices with different turn-on voltage but also for low-current, intermediate-current, and high-current regime (*e.g*. see Figures 2 (c-d)). In our case, the current regimes roughly correspond to TE, SCLC, and series resistance limited currents. In conventional diodes, the TE and SCLC regimes can be replaced with recombination and diffusive transport, respectively. The noise data for each regime has important implications for the device reliability assessment [48]. The low-frequency noise measured at low bias is sensitive to degradation of the active region [23,49]. At high bias, the measured noise reflects the degradation of the metal contacts and semiconductor layers contributing to the series resistance. Our data demonstrate that the difference in the noise level for different diodes is large at all currents (see Figure 3 (b)). The variations in the noise level are at their maximum, and span about three orders of magnitude, at the low currents. Interestingly, the smallest noise level was observed for the device with the low turn-on voltage and the highest noise was recorded for the devices with the high turn-on voltage. These observations attest to the potential of the noise spectroscopy for the diamond diode reliability assessment.

In conclusion, we reported the results of the investigation of excess noise in high-current diamond diodes. In high turn-on voltage diodes, the 1/$f$ noise dominated, which can be attributed to the higher concentration of traps responsible for noise in these diodes. The G-R noise was found to be characteristic for diamond diodes with lower turn-on voltages. The dependences of noise spectral density, $S_I$, on forward current show different slopes, which can be correlated with different transport regimes in the diodes. The characteristic time constants, extracted from the G-R noise data, reveal uniquely strong dependence on current, attributed to the specifics of the charge transport and recombination processes in our diamond diodes. The obtained results are important for developing of the noise spectroscopy-based approaches for the device reliability assessment for high-power diamond electronics.

**Acknowledgements**

The work at UCR and ASU was supported by ULTRA, an Energy Frontier Research Center (EFRC) funded by the U.S. Department of Energy, Office of Science, Basic Energy Sciences under Award #  DE-SC0021230. S.R. who contributed to noise data analysis acknowledges the support by CENTERA Laboratories in a frame of the International Research Agendas program for the Foundation for Polish Sciences co-financed by the European Union under the European Regional Development Fund (No. MAB/2018/9).

**Conflict of Interest**

The authors declare no conflict of interest.

**Author Contributions**

A.A.B. coordinated the project, lead data analysis and manuscript preparation. S.G. measured current-voltage and low-frequency noise characteristics; H.S. fabricated diamond diodes; F.A.K. contributed to device fabrication; R.N. supervised device fabrication; S.R., F.K., S.G. and R.N. contributed to the current-voltage and electronic noise data analysis. All authors contributed to the manuscript preparation.

**Supplemental Information**

The supplemental information with additional excess noise data and analysis is available at the Applied Physics Letters journal web-site for free of charge.

**The Data Availability Statement**

The data that support the findings of this study are available from the corresponding author upon reasonable request.

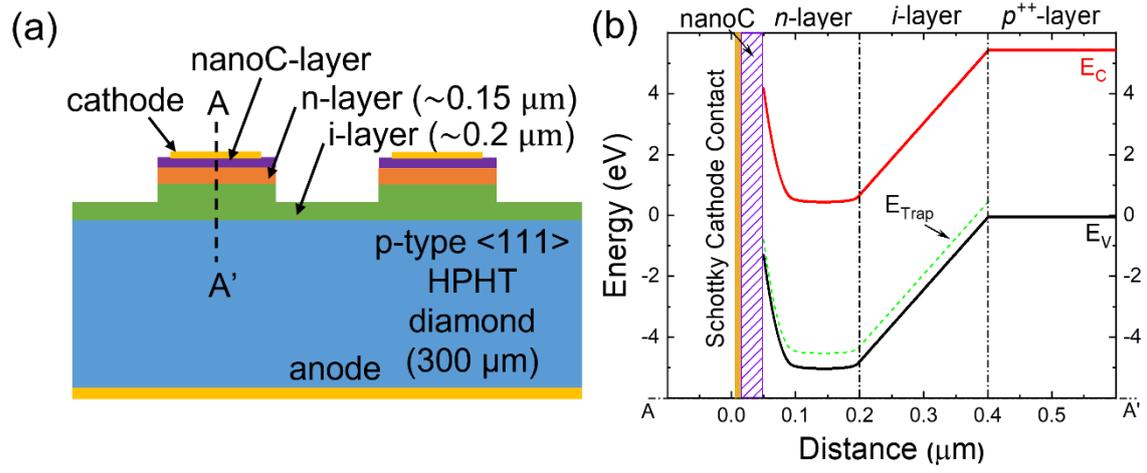

Figure 1

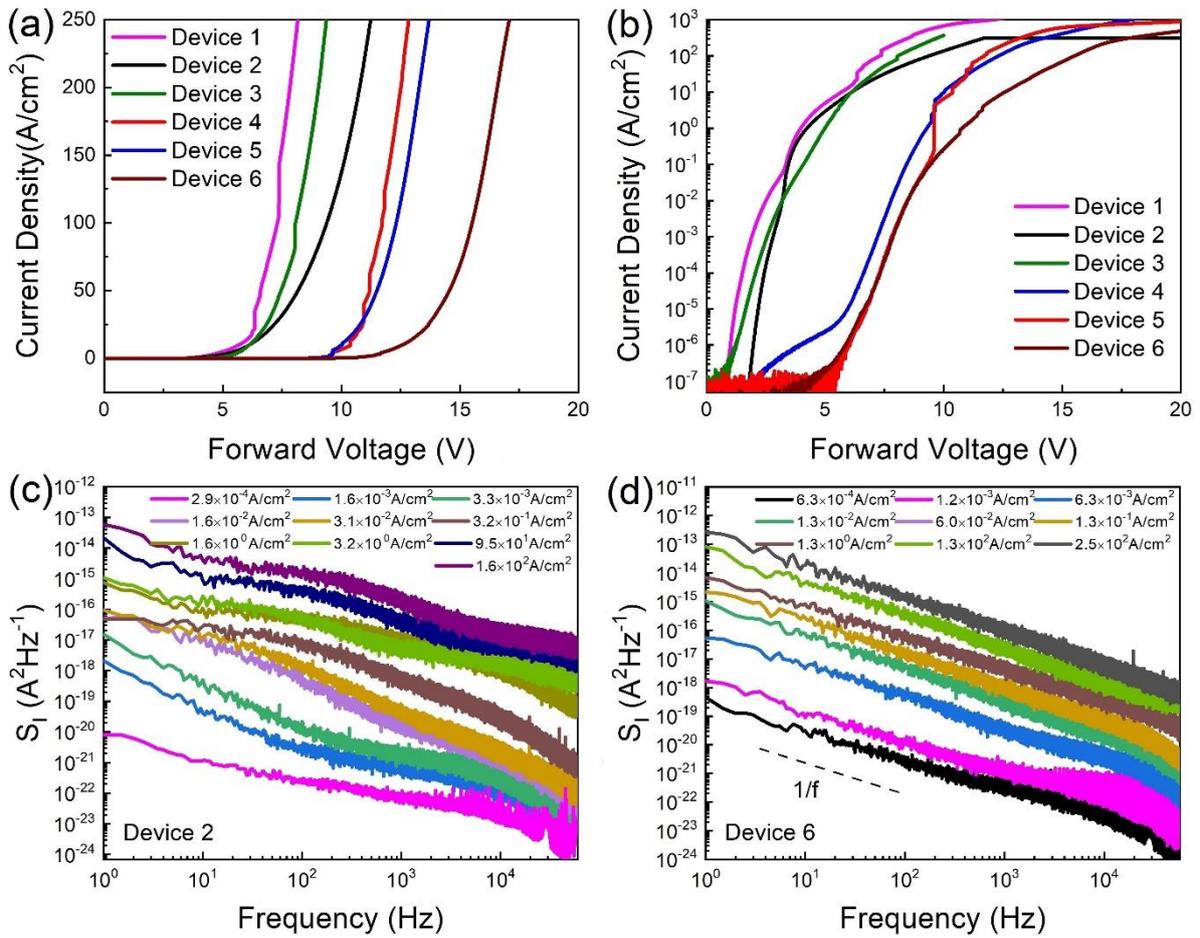

Figure 2

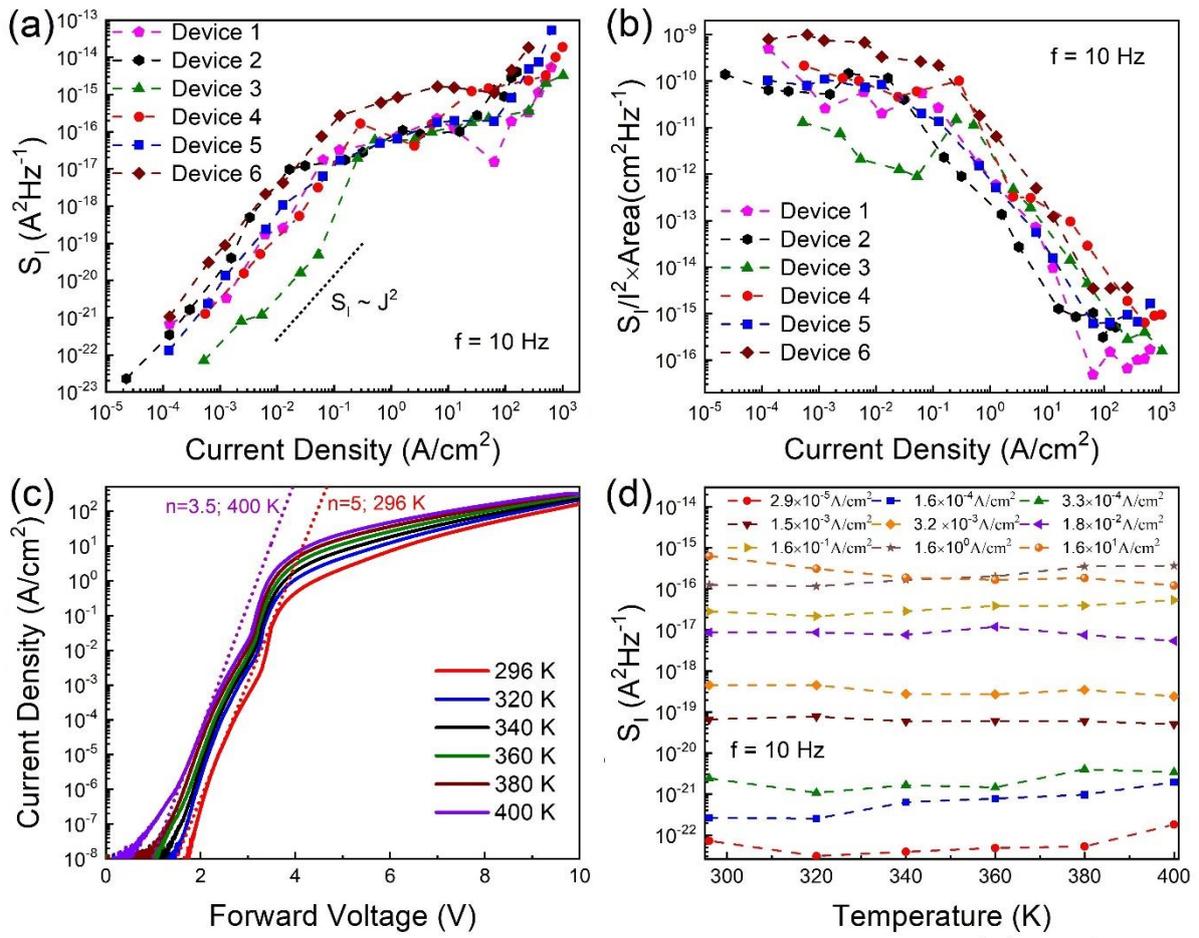

Figure 3

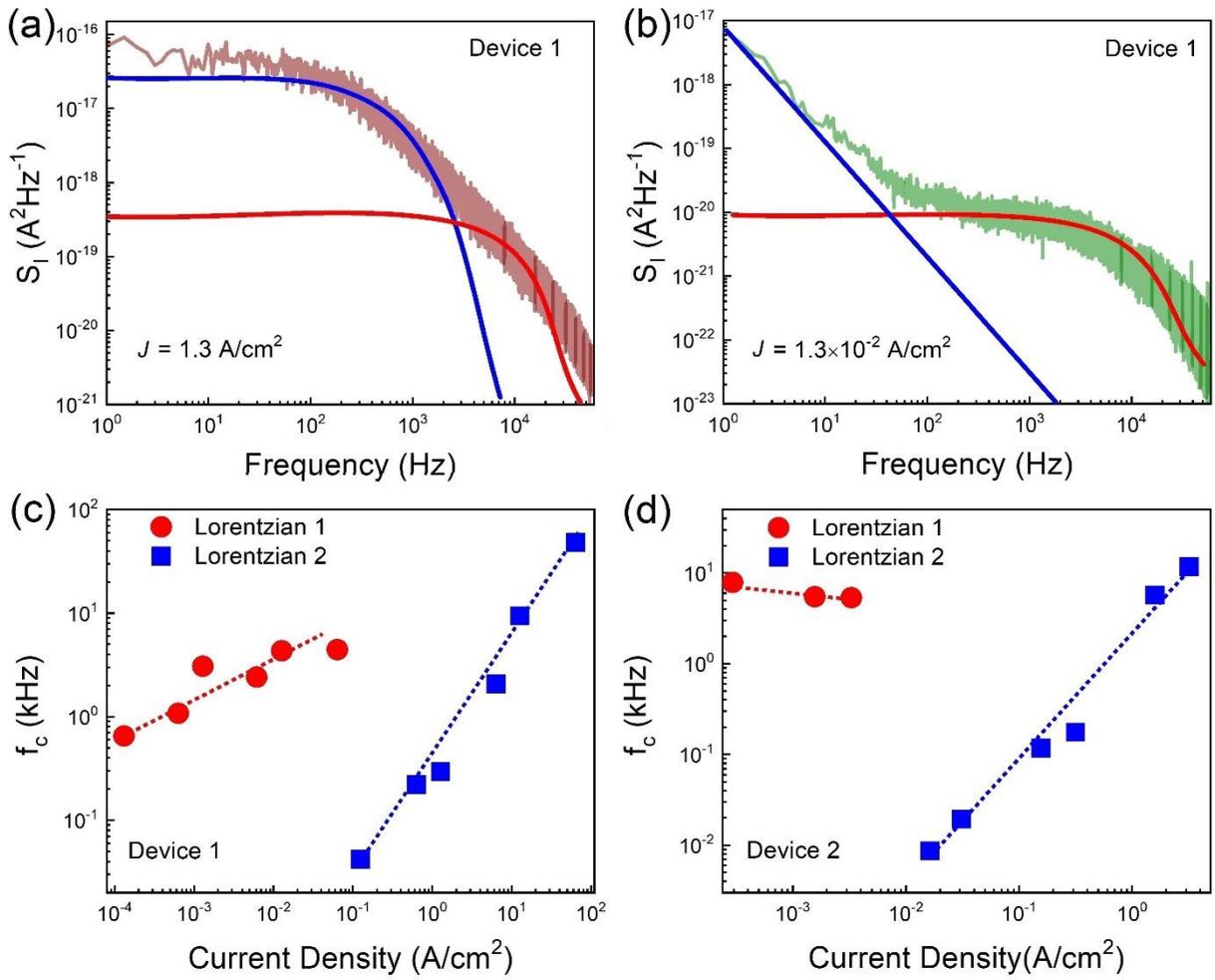

Figure 4